# Enhancing Educational Efficiency: Generative AI Chatbots and DevOps in Education 4.0

Edis Mekić, State University of Novi Pazar, Vuka Karadzica bb, 36300 Novi Pazar, The Republic of Serbia

Mihailo Jovanović, University of Criminal Investigation and Police Studies, 196 Cara Dušana Street, 11080 Belgrade-Zemun, The Republic of Serbia

Kristijan Kuk, University of Criminal Investigation and Police Studies, 196 Cara Dušana Street, 11080 Belgrade-Zemun, The Republic of Serbia

Bojan Prlinčević, Higher Technical Professional School in Zvečan, 6 Nusic str. 38227 Zvecan, The Republic of Serbia

Ana Savić, Academy Of Technical And Art Applied Studies Belgrade, 283 Vojvode Stepe 11010 Belgrade, The Republic of Serbia

**Abstract:** This research paper will bring forth the innovative pedagogical approach in computer science education, which uses a combination of methodologies borrowed from Artificial Intelligence (AI) and DevOps to enhance the learning experience in Content Management Systems (CMS) Development. It has been done over three academic years, comparing the traditional way of teaching with the lately introduced AI-supported techniques. This had three structured sprints, each one of them covering the major parts of the sprint: object-oriented PHP, theme development, and plugin development.

In each sprint, the student deals with part of the theoretical content and part of the practical task, using ChatGPT as an auxiliary tool. In that sprint, the model will provide solutions in code debugging and extensions of complex problems. The course includes practical examples like code replication with PHP, functionality expansion of the CMS, even development of custom plugins, and themes. The course practice includes versions' control with Git repositories.

Efficiency will touch the theme and plugin output rates during development and mobile/web application development. Comparative analysis indicates that there is a marked increase in efficiency and shows effectiveness with the proposed AI- and DevOps-supported methodology. The study is very informative since education in computer science and its landscape change embodies an emerging technology that could have transformation impacts on amplifying the potential for scalable and adaptive learning approaches.

## 1. Introduction

Digital transformation is a focal point of research across various academic disciplines. Digital transformation can be characterized as the process that encompasses altering the organization of activities by incorporating digital technologies and innovative organizational models to develop and implement innovative solutions [38]. The integration of these digital solutions establishes an environment where a deep understanding of theoretical concepts is not paramount for the workforce. Digital transformation introduces novel technologies, and skills, and opens up new job prospects. Rather than an in-depth grasp of theoretical concepts, employees require proficiency in the practical and interoperable utilization of emerging digital technologies [31].

We must be aware that new generations of students are digital native users, and we need to adapt their educational settings to fulfill their needs. This newly established reality must be understood to effectively build a new educational paradigm. For the new generation of students, we have to combine physical and virtual learning settings; students have to be capable of gathering knowledge from any place at any time; this brand new education form is known as Education 4.0. [19, 43]. One of the main concerns in engineering education is how to train future engineers to adapt to rapid technological changes. Future engineers must be trained for the jobs that do not exist right now [22]. With this in mind, it is more important than ever to train them with appropriate pedagogical strategies, such as experiment-based learning, research-based learning, and active learning [17]. Conversely, one of the primary requirements of emerging industries is the necessity for professionals with a broad array of learned skills. These skills encompass not only technical knowledge and proficiency in using new technologies but also soft (personal) competencies, a capacity for systems thinking, business acumen, technological literacy, and, notably, strong problem-solving capabilities [8].

Therefore, the subsequent research revolves around determining the learning methodological approach that aligns with Education 4.0 and can fulfill the essential learning requirements for future engineers in Industry 4.0. Given the high value placed on problem-solving capabilities, we require a methodology that facilitates the cultivation of this skill. Constructivism is a method deeply ingrained in a problem-solving approach to learning [11].

To understand why the constructivist methodology is in line with education 4.0 and industry 4.0 requirements, we must first comprehend fundamental constructivist notions. Constructivism is a fundamental pedagogical method that applies to active learning-based classrooms and emphasizes that students actively construct their knowledge, ideas, and concepts within a dynamic environment designed for teaching and learning practices [37]. The fundamental goal of constructivism is to allow students to construct their understanding. This idea emphasizes the significance of a learning environment that promotes active student participation, particularly through hands-on activities like making objects. Students participate actively in the learning process, taking actions that go beyond passive reception and encouraging problem-solving and critical thinking to gain a deeper comprehension of the subject matter [33].

While constructive added value to teaching and learning quality is thoroughly studied and proven, a further concern is the applicability of this methodology to current industrial needs. This methodology can be used to create new, innovative digital solutions during the educational cycle that adhere to industry best practices [40] and by applying appropriate laboratory and manufacturing practices [32]. These well-researched features of constructive methodology enabled straightforward application in Education 4.0 as baseline methodology, which encompasses another essential part of Education 4.0, project-based education [35,7,21].

As previously mentioned, to attain digital transformation and provide education in alignment with Education 4.0, we have opted for a constructive methodology. Another crucial aspect of this dynamic industry involves emerging tools with the capability to alter requirements, approaches, or production cycles rapidly. We will explore and incorporate two novel technologies to enhance the efficiency of training preparation and delivery. The first involves integrating AI Chatbots, and the second entails the utilization of DevOps tools.

The IT world has been profoundly influenced by these groundbreaking innovations since November 2022, when OpenAI debuted a sophisticated chatbot, ChatGPT 3.5. The chatbot in question was trained on a large dataset of web content [47]. The system's quality and variety have piqued the interest of educational institutions throughout the world, positioning it as a revolutionary instrument for education. At the same time, it has stimulated interest in the possible integration of AI solutions to modify and improve traditional teaching approaches [23].

The final piece of the puzzle for using new AI technologies, a constructive approach as part of digital transformation, and the advancement of an efficient education 4.0 environment is the application of proven tools that will aid in this process. Since digital transformation is the process of embracing new digital technologies that will revolutionize society and industry [16], we will employ IT industry technologies as instruments for raising awareness and developing new goods. Agile and DevOps techniques will be utilized to improve the efficiency of the overall learning process [6,29,18,9].

## 2. Research questions

As one aspect of our study, we will first use the constructive learning model as a basic methodological approach that matches the demands of Education 4.0. Then, as part of the constructive learning technique, we will extend this strategy to include AI chatbots and DevOps technologies and simultaneously implement them. Then, we will assess the degree of teaching efficiency before and after the adoption of AI chatbots and analyze the observed benefits of the latter.

Both issues will be addressed through the use of DevOps methods, which will help to improve the efficiency of the learning process in the context of digital transformation and Education 4.0.

These analyses will provide answers to the following research questions:.

In what ways may artificial intelligence chatbots be used to improve the efficiency of training planning and delivery in the context of Education 4.0?

In the context of digital transformation and Education 4.0, how might the use of DevOps technologies help to improve the efficiency of the learning process?

Answers to those questions will be acquired by proving the following hypotheses:

H1. Efficiencies of project-based learning are higher when using DevOps tools and AI chatbots in the constructive approach preparation phase.

H2. The efficiency of delivering learning material is higher.

## 3. Research desing

## 3.1 Constructive approach in line with Industry and Education 4.0

The initial step to implementing a constructive design in line with industry needs was to identify current practices in the existing industry. By analyzing these trends, we could determine how to define and modify the constructive approach. To achieve this, we reached out to four IT companies in Novi Pazar and distributed a brief questionnaire to 30 former students, now employed in the IT industry, through the university's alumni club. The questionnaire sought information on the types of modern managerial and DevOps technologies being utilized. The results revealed that, out of 250 projects over the past three years, 45% employed the Scrum methodology, 35% utilized Agile methodologies, and the remaining 20% followed various non-agile approaches. Given that Scrum is a type of Agile methodology, we chose Agile methodology as the baseline for implementing the constructive approach in this research.

Our analysis will focus on Agile-based methodology and the constructive approach. Drawing parallels between the two, we aim to propose modifications to the constructive approach, incorporating

Agile methodology. This adapted approach will form the basis for developing a methodological framework for learning that aligns with the requirements of Industry 4.0.

Agile is a comprehensive term encompassing various beliefs in software development. The primary aim behind introducing agile development was to establish a framework that enables development teams to incorporate changes into a project without undue effort. The fundamental values were advocated through the collaboration of 17 software engineering consultants who articulated the "Manifesto for Agile Software Development." This manifesto delineates values and principles for the agility of software development [4]. As agile constitutes a set of values, practices, and principles facilitating swift adaptation to changes in any aspect of software development, its implementation has led to cost reductions, enhanced productivity, improved quality, and increased satisfaction. Consequently, this paradigm has surpassed the IT industry [10].

All methodologies grounded in the Agile approach, including SCRUMM, Kanban, XP, and others [2], share a common foundation centered on implementing and adapting the product development cycle. The primary concept involves breaking down a project into smaller increments, which are subsequently divided into iterations for implementation. After each iteration, known as a sprint, a functional increment of the project is delivered. This iterative process encompasses distinct phases: planning, design, development, deployment, and review [35]. Simultaneously, the review phase serves as the entry point for planning the next iteration, as illustrated in Figure 1, encapsulating the entire iteration cycle.

Every successful Agile software development initiative commences with a planning stage. During this phase, the Agile product owner collaborates closely with stakeholders, the business team, developers, and future app users. Guided by the collective expertise of the team, the product owner shapes a comprehensive project vision by defining its purpose and goals, documenting business and user requirements, and prioritizing tasks while allocating resources effectively.

In all Agile-based methodologies, the design phase is a critical stage that follows the planning phase and precedes development. During this phase, the Agile team focuses on creating a blueprint for the software product, encompassing aspects such as architecture, user interface (UI), and user experience (UX). Design decisions are informed by the prioritized user stories and requirements established in the planning phase. The goal is to lay the groundwork for efficient and effective development by defining the overall structure of the software and ensuring alignment with the project vision. Iterative and collaborative, the design phase involves continuous feedback and adjustments, allowing for flexibility in accommodating changes as the development progresses. By emphasizing adaptability and responsiveness to evolving requirements, the design phase in Agile-based methodology facilitates a dynamic and iterative approach to software development.

Following the design phase, teams proceed to construct the initial iteration of the software. The development stage encompasses all production tasks in the Software Development Life Cycle (SDLC), including UX/UI design, architecture, and coding. Developing the first iteration often stands as the lengthiest phase within the Agile development lifecycle.

Upon completing the software development process, the Agile team deploys the application to the cloud or an on-premise server. With deployment, the product becomes live and accessible to customers, marking a significant milestone in the SDLC. This phase is typically celebrated as a noteworthy achievement. However, it signals the transition to the final stage.

Post-deployment, the work continues with ongoing maintenance to address bugs and uphold functionality. As users interact with the application, opportunities arise to collect valuable feedback, allowing for continuous improvement. This iterative feedback loop informs subsequent releases, contributing to the refinement and enhancement of the software in future iterations.

After defining the steps of the agile paradigm, it becomes crucial to integrate a constructive approach, a proven methodological strategy in engineering education, to seamlessly incorporate agile development. To prepare and equip future engineers for the challenges of emerging Industry 4.0, educational process must undergo a transformative process. This transformation involves the implementation of modern technologies that will shape an environment conducive to enhancing existing learning styles for both students and teachers. The amalgamation of established pedagogical methods, existing learning theories, and these new technologies is poised to generate novel approaches and concepts, fostering innovation in engineering education. This new concept in which Higher Education institutions apply new learning methods, innovative didactic and management tools, and smart and sustainable infrastructure mainly complemented by new and emerging ICTs to improve knowledge generation and information transfer processes is known as Education 4.0 [34].

The suggested Education 4.0 idea is built around four basic components. These components are:

(i) Competencies (the education and development of desirable essential skills in today's students),

(ii) Learning Methods (the integration of novel educational methods),

(iii) Information and Communication Technologies (ICTs) (the deployment of modern and emerging ICTs), and

(iv) Infrastructure (the utilization of cutting-edge equipment, amenities, and systems that enhance studying processes).

To implement Education 4.0 learning concepts in the context of Industry 4.0, particularly in an agile setting, a paradigm change is required—from an instructor-led knowledge production model to a participant-led and participant-centered knowledge creation model supported by collaboration. This transformation is at the core of the constructivist learning method [25,44,45]. A critical component is allowing students to create their own virtual environments and software applications, therefore broadening the resources accessible in the lab. This technique promotes learning via active engagement and hands-on experience [13]. The general structure for this technique is defined in three critical steps: planning implementation and learning given in Figure 2. [38]. The ideas of constructivist learning theory are very adaptable and may be easily integrated into a variety of processes and techniques by aligning certain stages of the theory with the proposed methods.

To achieve the aforementioned objectives, we proposed modification of a constructive framework for the implementation of agile development aligned with the needs of Industry 4.0 and Education 4.0. In the planning phase of this constructive framework, we will incorporate all fundamental components of Education 4.0. This involves defining the competencies that our courses will deliver, developing and proposing a learning method, and specifying and selecting ICT technologies for integration into the existing infrastructure.

The initial step in implementing this innovative approach is to define competencies needed from students, and this is studied in the following distinct courses on Content Management Systems within the Department of Technical Sciences, State University of Novi Pazar.

The curriculum for Content Management Systems aims to instill a comprehensive set of the following competencies:  developing a profound understanding of the theoretical foundations and practical applications of Content Management Systems (CMS). Students will acquire proficiency in designing, implementing, and maintaining content management solutions tailored to diverse organizational needs.

To enroll in the course for Content Management Systems, students are required to complete prerequisite courses in Web Design, Multimedia Systems, Internet Programming, and Software Engineering. This course carries six European Credits, as defined by the European Credit Transfer System (ECTS), involving 25 hours of student work throughout a 15-week semester. Since we want to apply agile methodology we divided the delivery of the course into three distinctive sprints every one lasting 5 weeks.

Throughout the entire semester, we have divided learning outcomes into three distinct segments, each lasting five weeks. These segments will specifically address the implementation phase of the constructive framework. Each part will focus on the general objectives outlined in the syllabus, and the implementation will follow the agile methodology through a structured sequence of three sprints.

The first objective is to acquire proficiency in advanced Object-oriented PHP practices, given that the majority of widely used Content Management Systems, such as WordPress, Joomla, Drupal, OctoberCMS, and OpenCart, are built on PHP. Notably, WordPress alone powers over 81 million websites, constituting approximately 43% of all websites on the internet, while PHP serves as the server-side scripting language for 77.5% of websites at the time of this research [49]. The second objective is to master the creation of custom themes for implementing various types of content in Content Management Systems (CMS). The third and final objective involves achieving expertise in fine-tuning the CMS and developing plugins to introduce new functionalities. These objectives collectively equip students with the knowledge needed to create sophisticated and modern web content, seamlessly integrating both front-end and back-end support. Every sprint will be modified, and based on the agile methodology framework provided earlier.

The initial stage, marking the commencement of all three sprints, is the planning phase. In this stage, the teacher, alongside assistants, articulates the purpose and objectives of the instructional subjects. Learning and teaching requirements must be documented, and tasks prioritized, with resources allocated accordingly. To maintain flexibility in this methodology, subsequent sprints must be devised by teachers based on the efficiency and outcomes achieved in the preceding sprint. These plans are detailed in the product backlog, featuring a list of learning items with assigned priorities. Estimates for each item, indicating the effort required for completion, and clear acceptance criteria for every item is also defined within this backlog.

To implement teaching with fundamental DevOps techniques, we employed Git version 2.43.0. Git, a distributed version control system, is designed to monitor changes in any collection of computer files, commonly employed for facilitating collaborative work among programmers engaged in concurrent source code development throughout the software development process. As of the completion of this research, Git commanded a substantial market share of 89.07% among DevOps tools [18]. To enhance visualization, we opted for the Sourcetree Git GUI, providing a user-friendly interface for effortless tracking of the proposed Git architecture.

At the beginning of each week, a set of training materials and solutions will be uploaded to Git for dissemination during the initial theoretical learning sessions. These solutions will be created using Large Language Learning Models such as ChatGPT as supporting tools. In the second part of learning, during the exercises, students will enhance and modify these solutions, employing the same supporting software solutions. After each week, a review will be conducted by the teaching staff to determine the content of the subsequent week's learning sprint.

To establish the necessary infrastructure, we opted for GitHub as the platform for repositories, serving as the storage for the actual codebase and its version history. The second facet of this infrastructure entails client machines, which serve as workstations for developers involved in code creation, modification, and testing. On each developer's machine, Git is installed, facilitating seamless interaction with the central repository. Developers utilize Git commands via either command-line interfaces or graphical user interface (GUI) tools to perform a spectrum of version control operations, including tasks such as cloning repositories, committing changes, branching, merging, and pushing/pulling code. Lastly, collaboration tools, encompassing issue trackers, project management tools, and communication platforms, collaborate with Git and establish a centralized hub for collaborative activities like discussion, issue tracking, and project planning, utilizing the Trello platform for this purpose.

Proof to our hypotheses will be based the implementation of engineering-based research [24]. Since this type of research should provide measurable results in the context of application in one specific instance of application of the tool in one distinctive event, while the proposed methodology can be used in a wider context of activities, we applied the case study as one of the engineering-based models applicable to these cases [5]. Also, we will present several code snippets developed and delivered by the teaching staff, which will be measures of the efficiency rate of our first hypothesis. Measurable results will also be delivered in the form of several projects or small snippets of code developed during learning classes and completed during the homework period for the second hypothesis.

## 4. Research implementation

### 4.1 Implementation of constructive approach in line with Industry 4.0 needs supported by agile methodological approach

To validate the previously mentioned hypotheses, we applied the proposed methodology within the Content Management Systems course, and we will provide a comprehensive explanation and showcase this specific case study.

### 4.2 Sprint 1

The primary goal of the initial sprint is to attain expertise in advanced Object-oriented PHP practices, given that the selected CMS, WordPress, is built and sustained using PHP. To initiate the planning for the first sprint, a meeting was conducted among the teaching staff. During this meeting, the overall objective of the first sprint was articulated, and key deliverables expected from this sprint were identified.

The overall objective is to equip learners with a robust foundation in PHP object-oriented programming, empowering them to design efficient and well-structured code.

Key Deliverables are to understand OOP principles, apply them in PHP, and creating basic PHP applications in object-oriented environment.

Next step in agile approach is to define objectives and topics which will be addressed in this sprint. The learning objective aims to furnish a thorough comprehension of PHP and its progression, with a specific focus on principles related to object-oriented programming. Commencing with an examination of PHP's design and administration, the curriculum advances to explore the evolutionary journey of PHP objects, spotlighting key milestones from PHP/FI to PHP 8.

Since the sprint was implemented over a five-week timeframe, the following subjects were covered each week.

The first week we covered essential object concepts like as classes, objects, attributes, and methods, as well as advanced features like static methods, abstract classes, interfaces, and traits. Following that, we covered fundamental tools for working with PHP objects, including as packages, namespaces, autoloading, and the Reflection API.

In the second week of the course, we focused on the integration of objects and design principles, highlighting code design, object-oriented vs. procedural programming, responsibility, cohesion, coupling, orthogonality, polymorphism, encapsulation, and class selection strategy.

The third week of the course focuses on giving a thorough grasp of design patterns and their applications in flexible and efficient object-oriented programming. We began by exploring design patterns and introducing its key features, such as organization, problem-solving technique, and the Gang of Four format. We also discussed the benefits of utilizing design patterns, highlighting their language independence, tried-and-true nature, collaborative design advantages, and broad acceptance by popular frameworks.

The forth week focuses on pattern concepts, including composition, inheritance, decoupling, and coding to an interface. We investigated certain design patterns for creating objects and enabling flexible object programming. These are the Singleton, Factory Method, Abstract Factory, Prototype, Service Locator, and Dependency Injection patterns. It also expands the learning to execute and express activities utilizing patterns such as Interpreter, Strategy, Observer, Visitor, Command, and Null Object.

The fifth week of sprint focused on the overarching goal of providing learners with the knowledge and abilities needed to effectively use design patterns in real-world object-oriented programming scenarios by evaluating previously completed tasks from the previous four weeks.

To complete the development agile phase, we provided the resources required, including the books and handbooks that served as the foundation for our course, online materials, and supplemental resources. During the initial sprint, we learn from Matt Zandstra's book, PHP 8 Objects, Patterns, and Practice: Mastering OO Enhancements, Design Patterns, and Essential Development Tools. Josh Pollock's "The Ultimate Guide To Object-Oriented PHP For WordPress Developers". Web resources were uploaded to the web repository, GitHub, while teachers stored local Git materials on their desktops. For all designated activities, we employed ChatGPT and taught it to grasp tasks created for students, as well as prepare exercises that were delivered to students as challenges.

Defined acceptance criteria were that students successfully recreate listings based on the List of requests which will be given at the beginning of the active part of the theoretical and practical part of learning classes.

The subsequent stage in agile development involves the preparation of learning materials and the establishment of a repository for these materials. Two crucial software tools will be employed: the first entails creating separate Git repositories on the computers of the teaching staff, while the second involves posting materials on an online repository. GitHub will be utilized for the online repository. Established in 2008, GitHub stands as one of the largest online platforms for software development and code hosting services, boasting over 100 million developers and 420 million repositories as of February 2024 [15]. This platform facilitates collaboration, with numerous developers and organizations contributing to repositories by introducing new features and resolving bugs [2]. Furthermore, developers glean insights from projects hosted on GitHub, incorporating lessons learned into their own endeavors [14]. It is also one of the most widely used platforms in the F/OSS (free and/or open-source software) space [40]. Also GitHub can implement coding as a 'social networked practice' [30], this is important for students in order to achieve cooperation and to be able to work in teams on the development of their tasks. This is one of the main prerequisites of constructive learning methodology.

Git, an open-source distributed version control and source code management system founded in 2005 by Linus Torvalds to facilitate cooperation on Linux operating system development, serves as the foundation for the GitHub platform. GitHub adds cloud-based hosting, graphical interfaces for browsers and desktops, and social networking features to Git. A GitHub repository has been established at https://github.com/ediemceach/MMSystemsLearningMaterial, and the initial framework of the Git structure for learning is now in progress. The first branch, named LearningMaterial, has been created on Git, where all necessary materials will be added as the learning process unfolds, as illustrated in Figure 4. Within this LearningMaterial branch, we will include all required materials and/or links to relevant literature or online resources. Subsequently, three consecutive branches will be created from this initial branch to correspond to the three planned sprints. The first sprint branch will be designated as OOPhP.

We will utilize this branch OOPhP to add step-by-step tasks and lessons for the first sprint. In addition to learning materials, we will use generative AI, namely ChatGPT, to improve learning efficiency.

Launched by OpenAI in 2018, ChatGPT is a variant of the Large Language Model (LLM) designed to emulate human language processing capabilities [42]. Utilizing advanced deep learning and sophisticated algorithms, ChatGPT can perform various language-related tasks, including text generation, question answering, and translation. It excels at comprehending context, enabling it to generate responses that closely resemble human language [28]. OpenAI introduced the ChatGPT-3.5 language model family in November 2022 and later released the ChatGPT-4 family in March 2023 [46]. This chatbot is proficient in engaging in coherent and contextually relevant conversations, offering responses based on its understanding of language and the context of the input prompts [12]. Accessible to anyone, ChatGPT allows users to sign up for the free conversational beta version of GPT-3.5 or subscribe to GPT-4 for a monthly fee of $20, and both versions can be used without requiring any training (OpenAI, 2023). After this we will engage in the final phase of development, that is backlog creation.

We identified tasks for each week of the sprint, breaking down the learning objectives into manageable units. Table 1 shows the list of tasks for Sprint 1. We also distributed tasks and separated them into smaller segments in Trello, as seen in Figures 5. This composition of tasks delivered in Trello is our sprint backlog.

Tasks will be executed following a standard approach to produce the theoretical segment of the learning material. Subsequently, we will train ChatGPT to replicate the previously provided solutions and utilize it as a supportive tool to devise more challenging assignments for students throughout the sprint. We aimed to allocate a one-week effort for each difficulty level outlined. Finally, clear criteria for approval have been defined for each task. All solutions and detailed implementations will be shared through the Git platform. This marks the deployment phase in the Agile methodology.

As an illustrative example, we will showcase the current implementation of a task from the second week of the sprint in Figure 6., indicating the branch where the analysis of our task will take place.

In the mentioned case, the task was to provide PHP code that defines classes for managing shop products, including ShopProduct, CdProduct, and BookProduct. It should also introduce an abstract class ShopProductWriter with a concrete implementation, Printer, for printing product details and summaries in various formats, including XML. Additionally, there's a function, fetchProductData, to retrieve product data from an SQLite database, and the code demonstrates creating instances of products and using the Printer class to output product information in XML.

Since we from the beginning used chatGPT as supporting tool we could recreate mentioned task with following set of prompts:

- "Generate a PHP code snippet illustrating the usage of the ShopProduct class with specific attributes and methods."
- "Explain the purpose and benefits of using an abstract class (ShopProductWriter) and its concrete implementation (Printer) in the provided code."
- "Describe the significance of the fetchProductData function and how it contributes to retrieving product information from a database."
- "Propose an extension to the existing shop product classes, such as adding a new property or method to enhance their functionality."
- "Provide an overview of how the Printer class generates XML output for shop products and explain its potential use cases in a real-world scenario."

We also used generative AI as support tool for delivery of the more complex example. This is proposed problem: Expand the given PHP code for the ShopProduct system by implementing additional functionalities and introducing a new product type. Specifically, introduce a DigitalProduct class that represents digital products (e.g., software downloads). Ensure that this class extends the ShopProduct class and adheres to the appropriate inheritance and encapsulation principles. Implement methods that are relevant to digital products, such as retrieving download links or license keys. Additionally, modify the ShopProductWriter class to accommodate the new DigitalProduct type and enhance the XML output to include information about digital products.

Those is set of prompts for achieving solution

- Prompt: Create a new PHP file (e.g., DigitalProduct.php) for the DigitalProduct class.
- Implement necessary methods in the DigitalProduct class, such as those related to download links or license keys.
- Modify the ShopProductWriter class to handle the new DigitalProduct type.
- Update the Printer class to accommodate the new DigitalProduct type in its print methods.
- In the main code, create an instance of DigitalProduct, add it to the Printer's products array, and generate the XML output.

It is important to notify that this solution is achieved by training chatGPT from the beginning of the sprint with proper data. Also several solutions received were debugged by the teaching staff and then Chabot was retrained in the line with solutions. Same approach was used for every task. last week of the sprint one complex project was given to all students to check did we managed to achieve all expected outcomes of learning. In the last week between the theoretical part of teaching problems were presented to students, up to the second part of learning 80% of students enrolled in the course managed to solve problems, remaining 20% solved problems as part of the homework assignment.

**4.3 Sprint 2**

Theme development is integral to Content Management Systems (CMS) due to its profound impact on the overall appearance and functionality of a website or application. Beyond mere aesthetics, themes play a pivotal role in shaping user experience and facilitating brand representation. They provide a platform for customization, allowing developers to tailor the visual elements such as layout, color schemes, and typography. The flexibility offered by themes ensures adaptability to specific needs and preferences, contributing to a positive user experience. Moreover, themes are designed to be responsive, ensuring optimal performance across various devices. Additionally, a well-structured theme positively influences search engine optimization (SEO) by contributing to clean code, fast loading times, and mobile responsiveness. As a result, theme development is not just about visual design but is a strategic aspect that influences user engagement, brand identity, and overall website functionality. Those reasons were the focus of the planning phase of the second sprint. At the teacher meeting, based on previous formulations, the following overall objectives and key deliverables were formulated.

The overall objective is to equip learners with a robust foundation for building custom themes that will be flexible, optimized, responsive, and custom-tailored.

Key deliverables will be delivered during five week sprint.

The first week of the course plan will introduce participants to the fundamentals of WordPress and theme building. This includes understanding how to install WordPress, configure critical settings, examine the structure of WordPress theme files, and get insight into content management via posts, pages, and custom post kinds.

During the second week, students will be delving further into the complex topics of theme structure and personalization. They will learn about theme style sheets, declarations, and how to use template files for headers and footers. The WordPress Loop, a key mechanism for content presentation, will be clearly addressed. In addition, the participants will experiment with page layouts to improve customization.

The next week will be focused on advanced theme creation methods. The idea and relevance of child themes will be discussed, with a focus on their function in expanding and modifying current themes. Participants will acquire hands-on experience by designing a basic child theme and learning how to apply it across several network sites. The week will also explore key aspects for topic selection and insights into theme frameworks.

During the last week, participants will create and build specialized themes. They will understand the value of rigorous preparation during the site idea and design stages. An introduction to theme frameworks will be offered, including their purpose and possible benefits. The attendees will also learn about licensing requirements and the importance of localization. They will also learn how to construct semi-static, media, magazine, and community themes, giving them a thorough grasp of theme creation in a variety of contexts.

The fifth week of the sprint focused on the overarching goal of providing learners with the knowledge and abilities needed to effectively use design to create their custom themes.

Now we returned to the GitHub repository that has been established at https://github.com/ediemceach/MMSystemsLearningMaterial, after completion of tasks in the git branch LearningMaterial/OOPhP we used the git command checkout to return to the LearningMaterial branch where we will create new sub branch ThemeDevelopment. Within this branch, we included all required materials and/or links to relevant literature or online resources.

Afterward, we generated a new backlog for the second sprint. As outlined in the proposed methodology, we identified tasks for each week of the sprint, breaking down the learning objectives into manageable units. Table 2 presents the list of tasks for Sprint 2, which was once again transferred to Trello to serve as our sprint backlog.

Tasks in the second week will be executed differently, given the theme development topic; WordPress provides a set of premade templates. In the first two weeks, we will analyze existing themes and utilize ChatGPT as a reverse engineering tool. The reverse engineering approach has already been proven effective in the development of new products in software engineering [1] and has also demonstrated its educational value [25].

As an illustrative example, we will cover the analysis of one of the premade templates. WordPress offers the following premade themes: Twenty Ten, Twenty Twenty-One, Twenty Twenty-Two, Twenty Twenty-Three, Twenty Twenty-Four. Before the analysis, we assigned a different theme to each of the 20 student groups.

The first step was to provide details about the structure of the folder containing the Twenty Twenty-One theme. This was achieved through two following prompts:

• We possess the following files in the WordPress directory: 404.php, archive.php, comments.php, footer.php, functions.php, header.php, image.php, index.php, package.txt, package-lock.txt, page.php, search.php, searchform.php, and style.css. Could you provide details about the functions of these files in the theme?

• Additionally, we also have the following folders in the theme: assets, classes, inc, and template part. Could you provide information on the contents of these folders?

ChatGPT successfully identified standardized files like footer.php, which contains the code for the footer section of your site, including copyright information, scripts, etc. While for some files, another deep analysis of code was needed to establish functionalities. ChatGPT identified them as probable functionality; for example: image.php: This file might be a custom template for displaying a single image attachment. While it was spot on in defining functionalities stored in separate folders.

One of the tasks was to provide a detailed analysis of the footer.php file. We provided 85 lines of code to ChatGPT to analyze, and he successfully divided the code into six important functionalities:

- **Main Content Closure:** The code closes the main content container (#content), which wraps the primary content area (#main).
- **Footer Widgets:** It includes a call to get_template_part for the 'footer-widgets' template part, suggesting that this theme supports footer widgets.
- **Footer Navigation:** If a footer navigation menu is defined, it displays it. The navigation is enclosed in a <nav> element with an aria-label for accessibility. It uses the wp_nav_menu function to output the menu items.
- **Site Info:** The footer contains information about the site, including the site name and logo (if custom logo is defined), a link to the front page if the site title is displayed, and a section for the privacy policy link if the function exists. Additionally, it displays a "Powered by WordPress" message with a link to the WordPress website.
- **WordPress Footer Hook:** The wp_footer() function is called, which is a WordPress hook that allows scripts and other code to be inserted just before the closing </body> tag.
- **HTML Closure:** The HTML document is closed with the closing </body> and </html> tags.

Finally, we prompted to identify and extract part of the code that manages footer navigation with the following prompt: Extract part of code that manages footer navigation. ChatGPT successfully delivered as shown in Figure 7.

Last week of the sprint, students developed their custom themes; 20 themes were successfully developed, and all students provided their repositories on GitHub for teaching staff to follow their iteration in development.

**4.3 Sprint 3**

Designer may add features and functionalities to a WordPress website using plugins that may not have been part of the initial installation. This is necessary if there is a need to modify the website to satisfy certain requirements. We will use plug-ins to implement two significant features throughout mentioned course.

The first is the application of a modular design strategy, which preserves customized elements in addition to fundamental functionalities. This concept is put into practice through the creation of plugins, which make site maintenance and upgrades easier without altering the site's fundamental design.

The second is expanding your website's functionality as it grows. This may be accomplished by using plugins, which provide a scalable solution that enables expansion of the functionality of a website without having to write a lot of code in the theme files.

Those are reasons why the overall objective of the third sprint iteration is to equip students with the skills and knowledge necessary for extending and refining the functionalities of CMS systems.

The same methodology as in the first two sprints is used, we defined key deliverables which will be delivered in five different week-long sprints. The first deliverable will be to understand the concept of WordPress plugins, explore available plugins, and grasp the advantages they offer.

The second key deliverable will be to comprehend the essential requirements for plugins, adhere to best practices, and gain proficiency in plugin development fundamentals, including plugin headers, paths, activation/deactivation functions, and uninstall methods.

The third key deliverable will be to learn to add menus and submenus, configure plugin settings using the Options API and Settings API, and maintain consistency in dashboard and settings management.

And fourth key deliverable will be to gain insights into WordPress security practices, understand user permissions, implement nonces, practice data validation and sanitization, format SQL statements securely, and optimize performance through caching and transients.

The fifth week of the sprint focused on the overarching goal of providing learners with the knowledge and abilities needed to effectively create their custom plugins.

Now we returned to the GitHub repository that has been established at https://github.com/ediemceach/MMSystemsLearningMaterial, after completion of tasks in the git branch LearningMaterial/OOPhP we used the git command checkout to return to the LearningMaterial branch where we will create new sub branch PluginDevelopment. Within this branch, we included all required materials and/or links to relevant literature or online resources.

Afterward, we generated a new backlog for the third sprint. As outlined in the proposed methodology, we identified tasks for each week of the sprint, breaking down the learning objectives into manageable units. Table 3 presents the list of tasks for Sprint 3, which was once again transferred to Trello to serve as our product backlog.

In this sprint, we will again use the same methodology as in the first sprint. We will present the theoretical part of the learning material to students, and then we will give a set of tasks to reproduce during learning and as homework. We will use ChatGPT as a supporting tool for students and teachers to jointly create new solutions. We will allocate a one-week effort for each difficulty level outlined. All solutions and detailed implementations will be shared through the Git platform.

**5. Research results, data analysis and study limitations**

In order to provide the results of the analysis of the proposed methodology, we must first define the group on which we delivered the newly established methodology. The course on the CMS system we have delivered to students since 2018. We employed a conventional teaching approach using Moodle-based course materials for students, up until the introduction of new AI and DevOps-supported techniques. In order to prove efficiency increase we will use following efficiency measurement formulas.

$$Efficiency\_Theme = Output\_Rate\_Theme \div Standard\_Output\ Rate\_Theme \times 100$$

$$Efficiency\_Plugin = Output\_Rate\_Plugin \div Standard\ Output\ Rate\_Plugin \times 100$$

$$Efficiency\_App = Output\_Rate\_App \div Standard\_Output\_Rate\_App \times 100$$

The most important value to define is the standard output rate, which is the rate of maximum performance or the maximum volume of work produced per unit of time using a standard method. This value will be defined in the case of the full implementation of AI and DevOps tools. Then we will measure the efficiency of the year when we did not use those tools.

As a starting point, we will use a group of 20 students during the 2023–2024 school year who took a new course. During these courses, every student completed four themes during the second sprint. Three themes were completed during the learning part, and one final theme was developed as a wrap-up during the last week of the second sprint. So we set Standard_Output_Rate_Theme=4, normalized per student.

The second important variable to measure efficacy is the standard output rate for plugin development. During these courses, 20 different plugins were developed during learning seasons, so Standard_Output_Rate_Plugin=20, normalized per student.

Finally, we need to define a standard output rate for the development of web and mobile applications using CMS as a framework. Since we have 10 completed applications and every team has two members, we will set Standard_Output_Rate_App=0.5 normalized per number of student-developed applications.

In this way, we set up the efficiency of the learning at the maximal value of 100%. To prove our hypotheses that the proposed approach indeed increased efficiency, we need to compare it with the efficiency for the periods of learning when we used the classical constructive approach without DevOps and AI technologies applied.

During this period, we used the classical approach to teaching, where we learned the theoretical basis for the topic and plugin development. Each topic covered 40% of the learning time during the semester. The rest of the semester was defined for the development of the fully functional CMS system and online and mobile applications, in which we used our CMS as a defined framework for application development. In those applications, we relied on the student's knowledge of PHP acquired in previous courses in web design and internet programming.

We covered the same topics covered in sprints two and three. Since we applied a constructive approach, teachers in that period delivered two WordPress themes and developed, during class 15, fully functional plug-ins. Students' obligatory activities were to develop a fully functional theme, one plug-in, and, as a final team effort, develop a fully functional CMS-based online and mobile application.

The first school year, which we will analyze, is 2019–2020. During this year, we had 12 enrolled students in the course. During the course, they developed 34 themes total, 190 plugins, and 3 applications in three teams with four members each. Those data gave us the following numerical values: Output_Rate_Theme=2,84; Output_Rate_Plugin=15,84; and Standard_Output_Rate_App=0.25. The calculation of efficiency gives us the following results: Efficiency_Theme_Dev=71%; Efficiency_Plugin=79.2%; Efficiency_App=50%. It is important to notify that course was held before official lockdowns because of Covid-19 pandemic, so these occurrence did not impact course significantly.

The second school year that we will analyze is 2020–2021, when we had 17 enrolled students. During the course, they developed 47 themes total, 267 plugins, and 5 applications in three teams with four members and one five-member team. Those data gave us the following numerical values: Output_Rate_Theme=2,76; Output_Rate_Plugin=15,70; and Standard_Output_Rate_App=0.29. The calculation of efficiency gives us the following results: Efficiency_Theme_Dev=69%; Efficiency_Plugin=78.5%; Efficiency_App=58%.

The third school year that we will analyze is 2021-2022, when we had nine enrolled students. During the course, they developed 26 themes total, 127 plugins, and 3 applications in three teams with three members. Those data gave us the following numerical values: Output_Rate_Theme=2.88; Output_Rate_Plugin=14.11; and Standard_Output_Rate_App=0.33. The calculation of efficiency gives us the following results: Efficiency_Theme_Dev=72%; Efficiency_Plugin=70.55%; Efficiency_App=66%.

Based on those results, we can see that the efficiency of the delivery of learning material and the efficiency of project-based learning are significantly higher when we employ AI and DevOps techniques in the learning curriculum

**Conclusion:**

In conclusion, this research introduces an innovative and holistic approach to computer science education by integrating Artificial Intelligence (AI) and DevOps methodologies into the learning process. The study focuses on Content Management Systems (CMS) development, emphasizing Object-Oriented PHP, Theme Development, and Plugin Development through structured sprints. The incorporation of real-world scenarios and the use of ChatGPT as a supportive tool contribute to a dynamic and practical learning environment.

The research demonstrates the effectiveness of the proposed methodology in enhancing students' skills and knowledge in CMS development. The structured sprints provide a systematic framework, allowing students to progress from fundamental concepts to advanced topics in a coherent manner. The use of AI, specifically ChatGPT, aids in code generation, debugging, and problem-solving, enriching the learning experience.

Furthermore, the integration of DevOps practices, including Git repositories for version control, aligns with industry standards. This prepares students for collaborative development environments and emphasizes the importance of versioning and code management—a crucial aspect of modern software development.

Efficiency measurement, employing quantitative metrics for theme and plugin development output rates, provides a clear assessment of the impact of the new methodology. Results indicate increased efficiency compared to traditional teaching methods, highlighting the potential of AI and DevOps integration in educational settings.

However, the research acknowledges certain challenges and limitations. The generalizability of findings beyond CMS development needs consideration, and the dependency on AI models introduces concerns related to biases and comprehension limitations. Resource intensity, student adaptation, and the long-term impact of the proposed methodology are areas that require ongoing evaluation and refinement.

In conclusion, this research sets the stage for redefining computer science education by embracing cutting-edge technologies. The positive outcomes observed in CMS development suggest the potential for broader application in diverse areas of computer science. Future work should focus on addressing identified limitations, exploring adaptability to different contexts, and ensuring sustained relevance in the rapidly evolving landscape of technology and education.

**Tables**

**Table 1. List of defined tasks for sprint 1.**

| Tasks for first Sprint | | | | |
|---|---|---|---|---|
| **I Week** | **II Week** | **III Week** | **IV Week** | **V Week** |
| **Classes and Object**<br>Setting Properties in a Class<br>Working with Methods<br>Creating a Constructor Method<br>Constructor Property Promotion<br>Default Arguments and Named Arguments | **Objects and Design**<br>Defining Code Design | **Patterns**<br>Name<br>The Problem<br>The Solution<br>Consequences | **Composition and Inheritance**<br>Composition and Inheritance<br>Using Composition | Use design patterns in real-world object-oriented programming scenarios by evaluating previously completed tasks from the previous four weeks |
| **Arguments and types**<br>Primitive types<br>Some Other Type-Checking Functions<br>Type Declarations:<br>Primitive Types<br>mixed Types<br>Union Types<br>Nullable Types | **Object-Oriented and Procedural Programming**<br>Responsibility<br>Cohesion<br>Coupling<br>Orthogonality | **The Gang of Four Format**<br>The Problem<br>The Solution | **Decoupling**<br>The Problem<br>Loosening Your Coupling | |
| **Inheritance problem**<br>Working with Inheritance<br>Public, Private, and Protected<br>Managing Access to Your Classes<br>Typed Properties | **Define classes**<br>Polymorphism<br>Encapsulation | **Why Use Design Patterns?**<br>A Design Pattern Defines a Problem<br>A Design Pattern Defines a Solution<br>Patterns Define a Vocabulary<br>Design Patterns Promote Good Design<br>Design Patterns Used by Popular Frameworks | **Why Use Design Patterns?**<br>Patterns for Generating Objects<br>Patterns for Organizing Objects and Classes<br>Task-Oriented Patterns<br>Enterprise Patterns<br>Database Patterns | |
| **Methods and Properties**<br>Static Methods and Properties<br>Constant Properties<br>Abstract Classes | **Four signposts**<br>Code Duplication<br>Constant Properties<br>The Jack of All Trades<br>Conditional Statements<br>Activity | | **The Singleton Pattern**<br>The Problem<br>Implementation<br>Consequences<br>**The Factory Pattern**<br>The Problem<br>Implementation<br>Consequences<br>**The Abstract Factory Pattern**<br>The Problem<br>Implementation<br>Consequences<br>**The Prototype**<br>The Problem<br>Implementation<br>Consequences | |
| **Traits**<br>Defining and Using a Trait<br>More Than One Trait<br>Combining Traits and Interfaces<br>Managing Method Name Conflicts with instead of<br>Aliasing Overridden Trait Methods<br>Using Static Methods in Traits<br>Accessing Host Class Properties<br>Defining Abstract Methods in Traits<br>Changing Access Rights to Trait Methods | **UML Diagrams**<br>Class Diagrams<br>Sequence Diagrams | | **The Interpreter Pattern**<br>The Problem<br>Implementation<br>Consequences<br>**The Strategy Pattern**<br>The Problem<br>Implementation<br>Consequences<br>**The Observer Pattern**<br>The Problem<br>Implementation<br>Consequences<br>**The Command Pattern**<br>The Problem<br>Implementation<br>Consequences<br>**The Null Object Pattern** | |

| | | | The Problem Implementation Consequences | |
|---|---|---|---|---|
| **Handling errors** Exceptions Final Classes and Methods The Internal Error Class Working with Interceptors Defining Destructor Methods | | | | |
| **PHP packages** PHP Packages and Namespaces Autoload | | | | |
| **The Class and Object** FunctionsChecklist Looking for Classes Learning About an Object or Class Getting a Fully Qualified String Reference to a Class Learning About Methods Learning About Properties Learning About Inheritance Method Invocation | | | | |
| **Reflections API** Getting Started Time to Roll Up Your Sleeves Examining a Class Examining Methods Examining Method Arguments Using the Reflection API | | | | |

**Table 2. List of defined tasks for sprint 2.**

| Tasks for second Sprint | | | | |
|---|---|---|---|---|
| I Week | II Week | III Week | IV Week | V Week |
| **Introduction to WordPress and Basic theme Concepts** Installing WordPress WordPress Settings Understanding WordPress Theme Files Introduction to Posts, Pages, and Custom Post Types | **Understanding Theme Structure and Customization** Theme Stylesheet and Declaration Template Files, Header, and Footer The WordPress Loop Working with Page Templates | **Using Child Themes** Concept and Importance of Child Themes Creating a Simple Child Theme Using Child Themes in Multiple Network Sites | **Planning the Theme** Site Concept and Design Introduction to Theme Frameworks Licensing and Localization | Custom theme creation |
| | Digging into the pre-prepared themes Working with the Premade Templates loop Using Template Tags and Conditional Content Enabling Features in functions.php Widgets, Custom Page Templates, and Menus | **Choosing a Theme** Criteria for Selecting Themes Theme Frameworks and Considerations | **Building Semi-Static, Media, Magazine, and Community Themes** Understanding different theme layouts Building sites for specific content types Extending functionality with BuddyPress | |

**Table 3. List of defined tasks for sprint 3.**

| Tasks for third Sprint | | | | |
|---|---|---|---|---|
| **I Week** | **II Week** | **III Week** | **IV Week** | **V Week** |
| **An Introduction to Plugins**<br>What is a plugin?<br>Available plugins<br>Advantages of plugins<br>Installing and managing plugins | **Security and Performance**<br>Security overview<br>User permissions<br>Nonces<br>Data validation and sanitization<br>Formatting sql statements<br>Security good habits<br>Performance overview<br>Caching<br>Transients | **Blocks and Gutenberg**<br>What is gutenberg?<br>Touring gutenberg<br>Practical examples<br>Technology stack of gutenberg<br>"hello world!" Block<br>Wp-cli scaffolding<br>Create-guten-block toolkit<br>Block directory | **Users and User Data**<br>Working with users<br>Roles and capabilities<br>Limiting access<br>Customizing roles | This structure ensures a comprehensive learning journey, starting with wordpress basics, progressing to theme development fundamentals, exploring advanced techniques, and concluding with the creation of specialized themes. Each week focuses on specific aspects, allowing for a gradual and in-depth understanding of wordpress theme development. |
| **Plugin framework**<br>Requirements for plugins<br>Best practices<br>Plugin header<br>Determining paths<br>Activate/deactivate functions<br>Uninstall methods<br>Coding standards | **Hooks**<br>Understanding hooks<br>Actions<br>Filters<br>Using hooks from within a class<br>Using hooks with anonymous functions<br>Creating custom hooks<br>Finding hooks | **Content**<br>Creating custom post types<br>Post metadata<br>Meta boxes<br>Creating custom taxonomies<br>Using custom taxonomies<br>A post type, post metadata, and taxonomy plugin | **Scheduled tasks**<br>What is cron?<br>Scheduling cron events<br>True cron<br>Practical use | |
| **Dashboard and Settings**<br>ADDING MENUS AND SUBMENUS<br>Plugin settings<br>The options api<br>The settings api<br>Keeping it consistent | **Javascript**<br>Registering scripts<br>Enqueueing scripts<br>Limiting scope<br>Localizing scripts<br>Inline scripts<br>Overview of bundled scripts<br>Polyfills<br>Your custom scripts<br>Jquery<br>Backbone/underscore<br>React | **Choosing a Theme**<br>Criteria for Selecting Themes<br>Theme Frameworks and Considerations | **REST API**<br>What the rest api is<br>What you can do with the rest api<br>Accessing the wordpress rest api<br>The http api<br>Wordpress' http functions<br>Bringing it all together | |

**Figures**

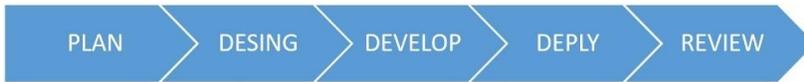

Figure 1. Agile methodology structure of single sprint

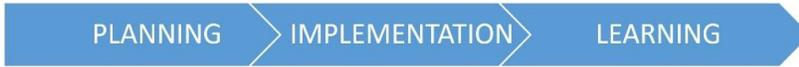

Figure 2. The general framework for constructive learning

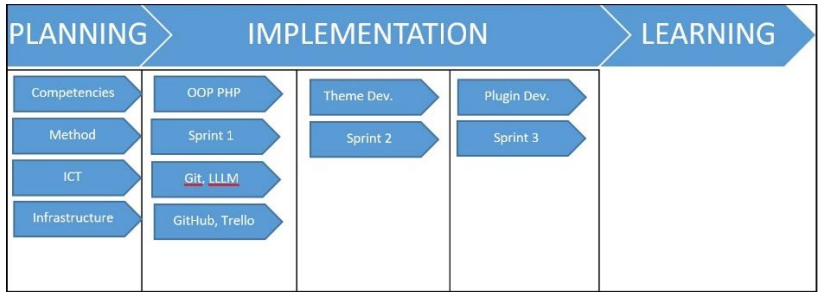

Figure 3. The general framework for constructive learning modification for seamless implementation of Agile-based methodologies

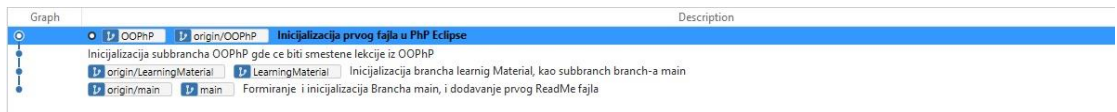

Figure 4. Fist sprint branch OOPhP in Git repository

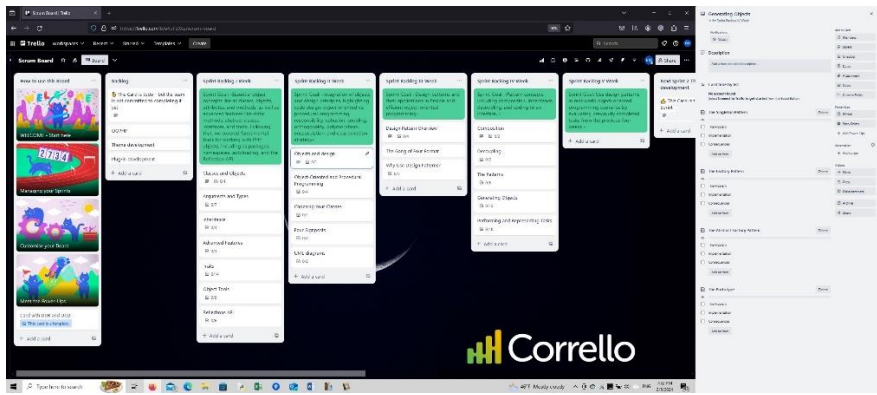

Figure 5. Product backlog in Trello

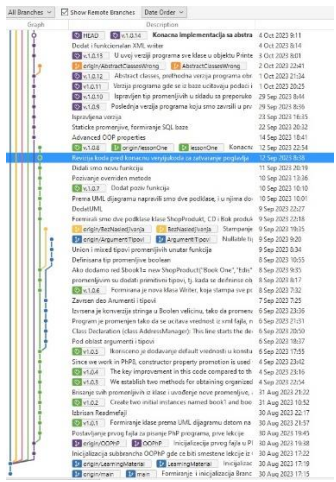

Figure 6. List of task in the Git platform

Figure 7. Code snippet that manage footer navigation extracted by ChatGPT